# Quantifying Uncertainties in the 2004 Sumatra-Andaman Earthquake Source Parameters by Stochastic Inversion

**Devaraj Gopinathan[1], Mamatha Venugopal[2], Debasish Roy[1], Kusala Rajendran[3], Serge Guillas[4], Frederic Dias[5]**

[1]Computational Mechanics Lab. Department of Civil Engineering. Indian Institute of Science. Bangalore 560012, India. [2]Department of Instrumentation and Applied Physics. Indian Institute of Science. Bangalore 560012, India. [3]Centre for Earth Sciences. Indian Institute of Science. Bangalore 560012, India. [4]Department of Statistical Science. University College London. London WC1E 6BT, United Kingdom. [5]School of Mathematics and Statistics. University College Dublin. Dublin 4, Ireland.

**Corresponding author: Kusala Rajendran (<kusala@ceas.iisc.ernet.in>)**

**Key Points:**

- Rational non-linear joint inversion of source slips and rupture velocities with rise times

- First reported inversion of rise times for a megathrust event

- Large parameter variances and skewness despite good waveform fit warrants better physical models

**Abstract**

Usual inversion for earthquake source parameters from tsunami wave data incorporates subjective elements. Noisy and possibly insufficient data also results in instability and non-uniqueness in most deterministic inversions. Here we employ the satellite altimetry data for the 2004 Sumatra-Andaman tsunami event to invert the source parameters. Using a finite fault model that represents the extent of rupture and the geometry of the trench, we perform a non-linear joint inversion of the slips, rupture velocities and rise times with minimal *a priori* constraints. Despite persistently good waveform fits, large variance and skewness in the joint parameter distribution constitute a remarkable feature of the inversion. These uncertainties suggest the need for objective inversion strategies that should incorporate more sophisticated physical models in order to significantly improve the performance of early warning systems.

**1 Introduction**

An uncommon opportunity for joint inversion of earthquake source parameters presents itself in the fortuitous detection, by satellites, of the tsunami [*Lay et al.,* 2005] caused by the 2004 Sumatra-Andaman event (2004 S-A), see Figure 1. The tsunami sea level anomaly (SLA) detected by Jason-1 and TOPEX/POSEIDON (T/P) altimeter-equipped satellites has no historical antecedent in its unambiguous demarcation of the signal and detection of its amplitude (~ 0.6 m). Not only is the SLA capture-duration commensurate with the entire rupture duration (~ 600 s), but the satellite track is also near-parallel to the entire rupture length (~ 1400 km). Given the rarity of tsunami detection by satellites, confluence of these remarkable features endows the SLA data with unprecedented value for inverting the source dynamics over the entire duration of the event. The corresponding uncertainty quantification affords a rare insight into the limitations in the underlying physical models.

Inversion of measured data for earthquake sources inevitably depends on the way the problem is posed and parameterized. Ill-conditioning and sensitivity issues may result from inaccurate physical models and inadequate or noisy measurements. Imposition of *a priori* subjective constraints (or regularization) is a common approach to computational amelioration of such drawbacks. However, the widely employed deterministic inversion schemes (*e.g.* the non-negative least squares method) cannot, by nature, comprehensively handle measurement noise, parameter uncertainties and multiple possible solutions. These methods do provide some insights into the possible solutions to a joint non-linear inversion, undertaken here for the slips, rupture velocities and rise times. However they cannot quantify the uncertainties about the solutions in terms of the ranges of likely values resulting from the inversion, or the nature of the distributions over these ranges - both necessary to assess the validity of scientific deductions that ensue from the inversion. While probabilistic methods endeavor to amend the pitfalls of the deterministic methods, many are ill-equipped for a global search, so a careful choice of the approach needs to be made when embarking on a probabilistic nonlinear inversion path. Finally, a single objective functional in the form of an error norm, employed in most of these schemes, may disregard local misfits and thus distort the statistical distribution of some of the inverted parameters whose influences might be purely local. There is thus a case for a stochastic and multi-objective global search *en route* to a more informative reconstruction.

The multi-objective evolutionary method [*Sarkar et al.,* 2015] employed here treats the measurements (Jason-1 and T/P SLA) and finite-fault source model parameters (slips, rupture velocities and rise times) as diffusive stochastic processes. Paucity of data and modeling deficiencies impart to the posterior probability distributions of the parameters a multimodal structure that reflects multiple solutions. The stochastic framework offers a natural and rational means to account for multiple solutions, hence simultaneously quantifying their uncertainties. Imposing minimal *a priori* constraints and thus according primacy to the measured data, our work showcases hitherto unreported large variances in most of the recovered parameters. This in turn calls for a fresh modeling approach by incorporating additional physics related to the dynamics of material discontinuities and heterogeneities such as large sediment cover.

**2 Posing the Forward Problem – Earthquake Slip to Seabed Uplift to SLA**

**2.1 Earthquake Slip to Seabed Uplift – Slip Parameterization by the Finite Fault (FF) Okada Model**

Information on the spatial distribution of coseismic displacements along the S-A arc is essential for computation of the consequent tsunami leading to an accurate description of the 2004 S-A event. The usual assumption that the final vertical seabed displacement can be equated with the initial sea surface change is invalid for slow ruptures, as reported for the Andaman segment [*Lay et al.,* 2005]. Okada [1985] presents explicit formulae for the spatial distribution of static displacement due to a slip across a rectangular fault patch in a semi-infinite isotropic homogenous medium. This requires nine parameters, *viz.* the patch length, width, depth, strike and dip angles, strike and dip slip components and geodetic coordinates. Many FF segmentations have been employed for this event [*Shearer and Bürgman,* 2010; *Poisson et al.* 2011]. We use the fourteen segments (or subfaults) from *Hirata et al.* [2006]. The static displacement obtained from Okada [1985] is multiplied by the trigonometric rise time function in Dutykh and Dias [2007] and Dutykh *et al.* [2013] to obtain the displacement history. Such mathematical separation of the temporal and spatial dependence of the slip history results in two additional parameters: slip initiation time (dependent on rupture velocity) and rise time. We presently invert slips, rupture velocities and rise times for all segments, fixing the remaining parameters as in *Hirata et al.* [2006].

**2.2 Seabed Uplift to SLA – Non-linear Shallow Water Equations (NSWE)**

Subduction zone earthquakes deform the seafloor triggering tsunamis. The displaced mass of water propagates as small waves in deep ocean but attains catastrophic amplitudes in shallow water due to amplification and coastal inundation. We have employed VOLNA [*Dutykh et al.,* 2011], a well validated finite volume NSWE solver. VOLNA efficiently handles complex processes such as dynamic seafloor elevation changes and simulates the entire tsunami life cycle of generation, propagation and inundation. A time series of deformation updates on the bathymetry generated at specified intervals forms the input to VOLNA. ETOPO1 bathymetry (1′-resolution) [*Amante and Eakins,* 2009] is utilized for the computational domain (8°S to 21°N and 82°E to 97°E), which contains the track portions of Jason-1 and T/P satellites where the tsunami was detected. The triangular mesh for the simulations is built upon the ETOPO1 grid. While reflections from the islands (north-east of the trench) could in principle contribute to the tsunami, the ~10 minute window of SLA capture is free of such subsequent reflections. Similar remarks hold for reflections from the regions adjacent to the Myanmar coast.

**3 Solving the Non-Linear Joint Inverse Problem – SLA to Source Slips, Rupture Velocities and Rise Times**

Using a linear shallow water equation (LSWE), the problem of inverting the slip alone from SLA (or tsunami wave height) is widely posed as linear [*Satake, 1*995; *Hirata et al.,* 2006; *Tanioka et al.,* 2006; *Fujii and Satake,* 2007]. In the LSWEs, the SLA and seabed uplift are linearly related, the latter in turn being linearly related to the slip via the Okada solution. This enables the employment of schemes like the non-negative least squares minimization [*Lawson and Hanson,* 1974]. Unfortunately, inversions for either rupture velocities or rise times are highly non-linear. In a linear setting, this amounts to fixing, perhaps arbitrarily, important parameters like those related to the FF array, rupture velocities, rise times and crustal property. Alternatively, these are deduced from *a priori* finite discrete sets. Complex interactions between the static (slip) and dynamic (rupture velocity and rise time) parameters cannot be reflected in such exercises. As a result, employing some form of variance reduction [*Hirata et al.,* 2006] or jackknifing [*Tanioka et al.,* 2006], cannot faithfully reproduce the statistical information and hence the uncertainties in the inverted parameters, see also *Beresnev* [2003] which discusses artifacts in the inverted solution in the context of seismic inversions. In fact, in a problem as complex as this, a significant part of the non-uniqueness could perhaps arise owing to the limitations and inaccuracies in the forward problem, *e. g.* the deformation model that should ideally account for thermo-visco-plasticity within a highly inhomogeneous and discontinuous medium. All the above factors highlight the importance of a non-linear, joint inversion of the source parameters attempted here.

**3.1 Non-linear Inversion Algorithm – Multi-objective Stochastic Evolutionary Optimization**

Amongst the possible options for joint inversion of slips, rupture velocities and rise times, stochastic methods perhaps offer the most rational framework for tackling the ubiquitous noisy character of real-world problems whilst naturally dealing with non-uniqueness in the inverted parameters. Surveys of inversion for the 2004 S-A event may be found in *Shearer and Bürgman,* [2010] and examination of various FF models in *Poisson et al.* [2011]. The relative unsuitability of a deterministic scheme for such problems arises in the possible ill-posedness, unphysicality or/and extreme sensitivity to regularization. Such regularizations may include imposition of constraints to limit the search space, slip-velocity positivity [*Olson and Apsel,* 1982], weak equalization of adjacent segments' slip [*Hartzell and Heaton,* 1983], uniform slips, average slip via seismic moment/magnitude ($M_0/M_W$) and others as discussed in *Beresnev* [2003]. Imposing such different constraints may well produce satisfying fits but giving totally different solutions. A significant step in the non-linear inverse problem approach was by Lorito *et al.* [2010], wherein the heat bath algorithm [*Rothman,* 1986] – a stochastic simulated annealing method – was used to invert for the multiple rupture velocities, rake angles and rigidities. However, it searched for the optimal solution only in discrete steps within bounded sets. Collapsing a continuum search to one on a discrete set may conceal important sources for non-uniqueness in the reconstructions associated with such non-linear inverse problems. Consequently, this discretization does not allow for proper quantification of uncertainties in the solution.

To precisely address these issues *en route* to a joint non-linear inversion, the stochastic optimization employed here is a variant of COMBEO – a Change of Measure Based Evolutionary Optimization [*Sarkar et al.,* 2014a, 2015; *Banerjee et al.,* 2009; *Teresa et al.,* 2014]. Its implementation is multi-objective. It aims to reduce the signed difference between the actual and computed SLA at each time instant wherever data are available (788 in total). This helps reveal a multi-modal structure in the posterior distribution, achieving a better fit for the whole duration of the waveform. COMBEO uses a filtered martingale approach for local optimization followed by a family of stochastic perturbation schemes, *e.g.* coalescence, scrambling and selection, for the global search. Fitted with a stochastic version of directional search that requires no derivative computation, COMBEO is usually found to approach the globally optimal solution with fewer iterations whilst retaining adequate flexibility for an exploration-exploitation trade-off. The need for incorporating *a priori* information/constraint is also appropriately met within the scheme. In the presence of multi-modality in the posterior distribution for the parameters (*i.e.* multiple possibilities for the solution), obtaining an acceptable solution must be based on a judicious combination of the local and global search features available in COMBEO. The generalized Bayesian search is based on Monte Carlo (MC) simulations and is done over a continuum of parameter intervals (see *Sarkar et al.* [2014b] for details of convergence of the stochastic search used in COMBEO).

The uncertainties in the forward model are accounted for by appropriate prior noise terms in the parameters evolutions, eventually reflected in their joint posterior distribution. The search provides a natural means to empirical quantifications of the various uncertainties involved. The number of particles (ensemble size) was limited to five since it provided satisfying results with no need to increase further. The uncertainty in the measurement is another important, yet often overlooked factor, which is to an extent accounted for in COMBEO by the measurement noise. However, being an MC scheme, implementing COMBEO is a computationally intensive endeavor constituting of repeated numerical evaluations of VOLNA.

**3.2 Measurement Data – Uncommon SLA for the 2004 S-A Tsunami**

Four altimeter equipped satellites, *viz.* Jason-1, T/P, ENVISAT and GFO mapped the ocean following the 2004 S-A event. SLA levels of considerable amplitudes were recovered from the signals after post-processing. Jason-1 (456 measurements) and T/P (332 measurements) SLA data has been used since these satellites captured the leading wave front of the tsunami in the deep Bay of Bengal region. The early onset time of anomaly detection (~2 h post rupture initiation) and the location of the satellite passes in deeper parts of Bay of Bengal are crucial not only for detecting the primary wave front but also for excluding coastal reflections and dispersive effects, thus precluding the need for shallow water/coastal bathymetry data and dispersive wave propagation models. The altimeter accuracy (3 cm), temporal (~1 s) and spatial (27 km) along-track resolution and quality post-processing of altimetry data are other vital features favoring the use of SLA data for inversion. The multisatellite time-spatial interpolation (MSTSI) method [*Hayashi,* 2008] is used for additional post-processing to extract the tsunami signal (Figure S1 in SI). Neither are additional points added to the SLA data via interpolation (in order to give due primacy to measured data) nor is any relative weighing factor used between the satellite data.

### 3.3 Recovered Dynamic Source Parameters

Here we report three cases of inversions: (A) inversion of slips alone, (B) joint inversion of slips and rupture velocities and (C) joint inversion of slips, rupture velocities and rise times. For A, rupture velocities of 2.5 km/s and rise times of 60 s are assumed with initial slips of 10 m. For B, rise times of 60 s are assumed with initial slips and rupture velocities of 10m and 2.5 km/s respectively and for C, the initial values are 10 m, 2.5 km/s and 60 s. Minimal *a priori* constraints are imposed in the form of bounds for the evolving parameters: 0 – 100 m for slips, 0.5 – 4 km/s for rupture velocities and 30 – 300 s for rise times.

In all the three cases, the inverted parameters are reported after the $l_2$-norm of the multi-objective error attains steady-state at around thirty iterations, beyond which only very small fluctuations are noticed (Figure 2a). Similar observations apply to the moment magnitudes and seismic moments which attain steady-state at about 9.35 and 13.5 x $10^{22}$ Nm respectively (Figures 2b and S2 in SI). The effect of realizing a steady-state in the error norm is clearer in the corresponding fits between measured SLA and computed waveforms for both the satellites (Figures 3 and S3 in SI). The waveforms computed after thirty iterations are seen clustered about the measured SLA for all the three cases. Careful observation reveals that the computed waveforms tend to match better the measured SLA waveform over certain intervals on inclusion of more parameters for inversion. Thus relatively speaking, the envelope of computed waveforms for C (inversion of slips, rupture velocities and rise times) enables a better reconstruction than B (inversion of slips and rupture velocities), which in turn provides a better fit than A (inversion of slips alone).

**4 Present Results in the Context of Previously Reported Inversions**

In the absence of previously reported exercises on similar lines, a comparative assessment of the iterative evolution of parameters, errors and other derived quantities (like $M_W$ and $M_0$) reported here is not possible. The fast and near-monotonic decay of the error norm results from the superior form of the exploration-exploitation tradeoff embedded in COMBEO and owing to a multi-objective search. Variances in the parameters captured by the method in all the three cases are quite significant even though the respective errors have attained a steady state (Figure 4). The variances are also different for each case.

For none of the cases the slips reach the prescribed upper bound of 100 m. High mean slips of 30 – 60 m are consistently observed in segments 4 and 5. The variances are visibly quite different between the three cases, especially in segments 7 – 9. Mean slips of 20 – 30 m are also consistently seen in the northernmost segments (12 – 14) where the rupture terminates. Although unusually high, they are supported by a good fit in the northern end of the waveforms. Interestingly, coseismic land level changes observed both from geodetic and field evidence also suggest variations along the rupture zone with its northern terminus (segment 14) showing relatively larger vertical offset [*Rajendran et al*, 2007]. Overall, the slips here are relatively higher than previously reported, *e.g.* Table 2 in *Shearer and Bürgman* [2010] and in *Poisson et al.* [2011]. This also results in the marginally higher values of $M_0$ than that reported in the former (5.7 – 11 x $10^{22}$ Nm). A close match could have been enforced, *e.g.* by assuming a steeper dip angle in the FF geometry [*Shearer and Bürgman,* 2010], a strategy avoided here. Similarly, the terminal high slips could also have been obviated by imposing the *a priori* constraint of zero slip in the rupture termination zone. A closer examination of the SLA reveals unmatched fluctuations that could be either ascribed to coarse segmentation or deficiencies in the model.

Large variances are also seen in the rupture velocities although they are curtailed by the prescribed upper bound of 4 km/s. Significant is the discrepancy when the rise times are included in the inversion, particularly for segments 7 – 9 and the last two terminal segments (13 and 14). The variances nearly encompass the range of values given in Table 2 and 3 of *Shearer and Bürgman* [2010] *i.e.* 1.5 – 4.1 km/s. To the best of the author's knowledge, there are no existing inversions for rise times using SLA data let alone joint inversions. The variances in rise times seem relatively larger for segments 4, 5, 8, 9 and 11 – 13.

Parameter distributions in many segments are clearly asymmetric, highlighting the non-Gaussianity of the posterior distribution. This is manifested for instance in segment 7 of Figure 4a, wherein the nature of skewness changes markedly in the three different inversion cases. Moreover, considerable variances in parameters found here explain the plethora of related values reported previously, clearly mandating a more nuanced interpretation of the results than is typically done.

**5 Discussions and Conclusions**

The rational approach employed in COMBEO, the evolutionary stochastic optimization scheme adopted here, *en route* to a joint non-linear inversion while also addressing the problems inherent in common deterministic schemes constitutes a significant first step forward for earthquake source recovery problems. The rational yet flexible machinery of advanced non-linear inversion techniques allows for an attempt not only at recovering the dynamic source model of the 2004 S-A event but also at accounting for complex interactions between the static (slip) and dynamic (rupture velocity and rise time) parameters. Instead of usual variance reduction by incorporating arbitrary *a priori* constraints, the possibility of multiple solutions is retained by imposing a minimal set of constraints. The resulting uncertainties could originate from the limitations in the physical models and paucity of measurement data. In view of the large variances observed, this study emphasizes the need for better physical models.

An important component in modeling uncertainties emerges from the simplifying assumptions in the Okada model, *e.g.* isotropy, homogeneity and elastic half space. The lateral variations in the material property and in the bathymetry are also unaccounted for. Similarly important is the choice of parameterization in the practical utilization of the Okada model which includes the geometry of the FF array defined by its orientation, spatial spread, number of segments and grid spacing [*Poisson et al.*, 2011]. The 2004 rupture lies in the Bay of Bengal, where the ocean floor is overlain by kilometers thick sediment cover [*Whittaker et al.,* 2013] especially near its northern termination where it is ~5 – 10 km thick. The non-linear and non-classical behavior [*Beresnev and Wen,* 1996; *Johnson,* 2006] of this compacted granular media is not reflected in the Okada model. In the absence of such physical models, incorporating spatial distribution of material properties in the finite element method (FEM) [*Masterlark and Hughes,* 2008; *Dutykh and Dias,* 2010] only addresses the problem in a limited manner. COMBEO's stochastic compensation for this unaccounted material behavior may be a reason for (i) the relatively high mean slips throughout the rupture, (ii) the apparently discrepant high slips (40 – 60 m) in segment 5 and (iii) the consistently recovered high mean slips (20 – 30 m) in the rupture termination region.

Another source of epistemic uncertainty enters the joint inversion due to the paucity of dynamic source (*i.e.* slip history) models [*Dutykh and Dias,* 2007]. Usual inversions, where slips alone are inverted, fix the slip initiation and rise time of the segment, thus making the segment slip the only unknown in its slip history. The amelioration attempted here is by allowing joint evolution of the segment slip, its initiation time (deduced from its rupture velocity) and rise time. Parameterization of rupture dynamics using a rupture velocity and rise time per segment invariably makes them sensitive to the segment geometry, besides the coarse resolution resulting from segment-wise rise times given the infinite frequency spectrum of the actual slip history. Accordingly, our parameterization is a significant development, though not the best possible. As with slips, this uncertainty is stochastically captured and quantified by COMBEO. Consequently, it is pertinent not to lose sight of inadequacies in the existing physical models and lack of suitable source models while interpreting any inversion result, including the findings here. Finally, it is remarkable that although uncertainties in measurement (SLA data) also contribute to the variances, the data suffices to constrain the parameters within the bands reported. This brings to focus the practice adopted here of according due primacy to measurement data over *a priori* constraints. As additional data is incorporated within the scheme, a reduction in parameter variances ought to be indicative of the quantum of additional information contained therein.

Although previous works point toward high slips in the northern regions of the rupture, the present study offers a rational confirmation. This is due to the consistent recovery of high slips in the vicinity of the rupture termination in all three inversion cases. Some possible causes include non-linear wave amplification through sediments [*Beresnev and Wen,*1996] and change of material properties within the subduction zone. This again emphasizes the need for fresh models incorporating additional physics pertaining to the rupture dynamics and consequent deformation propagation.

The results here can be improved by incorporating various enhancements *e.g.* FEM modeling of the source, adopting the spherical co-ordinate version of the NSWE, high resolution bathymetry datasets *etc*. Likely improvements withal, parameter reconstruction for a problem as complex as this would continue to exhibit a significant uncertainty band. This shows that a stochastic scheme like COMBEO would continue to remain relevant as an effective tool for tsunami source recovery. The characteristics of the uncertainties reported here (*e.g.* skewness, variance, complex inter-dependencies) strongly suggest the need for developing more sophisticated physical models. Consequently, the performance of early warning systems could be dramatically improved by a rational inclusion of uncertainties in the prediction of hazard footprints.

**Acknowledgments and Data**

SLA data is from http://rads.tudelft.nl/rads/rads.shtml using default post-processing. G.D., M.V. and D.R. acknowledge the **D**efence **R**esearch **D**evelopment **O**rganization (Grant#

ERIP/ER/1201130/M/01/1509). D.R. and K.R. acknowledge the **I**ndian **N**ational **C**oastal **I**nformation **S**ervices (Grant# INCOIS:F&A:D3:012:2011) for part of this work. D.R., K.R. and S.G. acknowledge the UK **D**epartment **f**or **I**nternational **D**evelopment, the **N**atural **E**nvironment **R**esearch **C**ouncil and the **E**conomics and **S**ocial **S**cience **R**esearch **C**ouncil as part of the Science for Humanitarian Emergencies and Resilience Programme (Big Data for Resilience Call) for part of this work. S.G. acknowledges the NERC research programme **P**robability, **U**ncertainty and **R**isk in the **E**nvironment (Grant# NE/J017434/1). F.D. and S.G. acknowledge the European Union's Seventh Framework Programme for research, technological development and demonstration under grant agreement ASTARTE# 603839.

**References**

Amante, C., and B. W. Eakins (2009), ETOPO1 1 arc-minute global relief model: Procedures, Data Sources and Analysis, NOAA Technical Memorandum *NESDIS NGDC-24*, National Geophysical Data Center, Marine Geology and Geophysics Division, Boulder, Colorado.

Banerjee, B., D. Roy, and R. M. Vasu (2009), A pseudo-dynamic sub-optimal filter for elastography under static loading and measurements, Physics in medicine and biology, 54(2), 285-305, doi: 10.1088/0031-9155/54/2/008.

Beresnev, I. A. (2003), Uncertainties in finite-fault slip inversions: to what extent to believe? (A critical review). Bull. Seismol. Soc. Am., 93(6), 2445-2458.

Beresnev, I. A., and K.-L. Wen (1996), Nonlinear Soil Response–A Reality?, Bull. Seismol. Soc. Am., 86(6), 1964-1978.

Dutykh, D., and F. Dias (2007), Water waves generated by a moving bottom, in Tsunami and Nonlinear waves, 65-95, Springer, Berlin Heidelberg, doi:10.1007/978-3-540-71256-5_4.

Dutykh, D., and F. Dias (2010), Influence of sedimentary layering on tsunami generation, Computer Methods in Applied Mechanics and Engineering, 199(21-22), 1268-1275, doi:10.1016/j.cma.2009.07.011.

Dutykh, D., R. Poncet, and F. Dias (2011), The VOLNA code for the numerical modeling of tsunami waves: Generation, propagation and inundation, European Journal of Mechanics-B/Fluids, 30(6), 598-615, doi:10.1016/j.euromechflu.2011.05.005.

Dutykh, D., D. Mitsotakis, X. Gardeil, and F. Dias (2013), On the use of the finite fault solution for tsunami generation problems, Theoretical and Computational Fluid Dynamics, 27(1-2), 177-199, doi:10.1007/s00162-011-0252-8.

Fujii, Y., and K. Satake (2007), Tsunami source of the 2004 Sumatra–Andaman earthquake inferred from tide gauge and satellite data. Bull. Seismol. Soc. Am., 97(1A), S192–S207, doi:10.1785/0120050613.

Hartzell, S. H., and T. H. Heaton (1983), Inversion of strong ground motion and teleseismic waveform data for the fault rupture history of the 1979 Imperial Valley, California, earthquake, Bull. Seismol. Soc. Am., 73(6A), 1553-1583.

Hayashi, Y. (2008), Extracting the 2004 Indian Ocean tsunami signals from sea surface height data observed by satellite altimetry, Journal of Geophysical Research: Oceans, 113(C1), C01001, doi:10.1029/2007JC004177.

Hirata, K., K. Satake, Y. Tanioka, T. Kuragano, Y. Hasegawa, Y. Hayashi, and N. Hamada (2006), The 2004 Indian Ocean tsunami: Tsunami source model from satellite altimetry, Earth, Planets and Space, 58(2), 195-201, doi:10.1186/BF03353378.

Johnson, P. A. (2006), Nonequilibrium nonlinear dynamics in solids: State of the art, in Universality of Nonclassical Nonlinearity, 49-69, Springer, New York, doi:10.1007/978-0-387-35851-2_4.

Lawson, C. L., and R. J. Hanson (1974), Solving least squares problems, Prentice-Hall, Englewood Cliffs.

Lay, T., et al. (2005), The great Sumatra-Andaman earthquake of 26 December 2004, Science, 308(5725), 1127–1133, doi:10.1126/science.1112250.

Lorito, S., A. Piatanesi, V. Cannelli, F. Romano, and D. Melini (2010), Kinematics and source zone properties of the 2004 Sumatra-Andaman earthquake and tsunami: Nonlinear joint inversion of tide gauge, satellite altimetry, and GPS data, Journal of Geophysical Research: Solid Earth, 115(B2), B02304, doi:10.1029/2008JB005974.

Masterlark, T., and K. L. Hughes (2008), Next generation of deformation models for the 2004 M9 Sumatra-Andaman earthquake. Geophysical Research Letters, 35(19), L19310, doi:10.1029/2008GL035198.

Okada, Y. (1985), Surface deformation due to shear and tensile faults in a half-space, Bull. Seismol. Soc. Am., 75(4), 1135-1154.

Olson, A. H., and R. J. Apsel (1982), Finite faults and inverse theory with applications to the 1979 Imperial Valley earthquake. Bull. Seismol. Soc. Am., 72(6A), 1969-2001.

Poisson, B., C. Oliveros, and R. Pedreros (2011), Is there a best source model of the Sumatra 2004 earthquake for simulating the consecutive tsunami?, Geophysical Journal International, 185(3), 1365-1378, doi:10.1111/j.1365-246X.2011.05009.x.

Rajendran, C. P., K. Rajendran, R. Anu, A. Earnest, T. Machado, P. M. Mohan, and J. Freymueller (2007), Crustal Deformation and Seismic History Associated with the 2004 Indian Ocean Earthquake: A Perspective from the Andaman–Nicobar Islands, Bull. Seismol. Soc. Am., 97(1A), S174-S191, doi:10.1785/0120050630.

Rothman, D. H. (1986), Automatic estimation of large residual statics corrections. Geophysics, 51(2), 332-346, doi:10.1190/1.1442092.

Sarkar, S., D. Roy, and R. M. Vasu (2014a), A perturbed martingale approach to global optimization, Physics Letters A, 378(38), 2831-2844, doi:10.1016/j.physleta.2014.07.044.

Sarkar, S., S. R. Chowdhury, M. Venugopal, R. M. Vasu, and D. Roy (2014b), A Kushner–Stratonovich Monte Carlo filter applied to nonlinear dynamical system identification, Physica D: Nonlinear Phenomena, 270, 46-59, doi:10.1016/j.physd.2013.12.007.

Sarkar, S., D. Roy, and R. M. Vasu (2015), A global optimization paradigm based on change of measures, R. Soc. open sci., 2(7), doi:10.1098/rsos.150123.

Satake, K. (1995), Linear and nonlinear computations of the 1992 Nicaragua earthquake tsunami, Pure Appl. Geophys., 144(3/4), 455-470, doi:10.1007/978-3-0348-7279-9_6.

Shearer, P., and R. Bürgmann (2010), Lessons learned from the 2004 Sumatra-Andaman megathrust rupture, Annual Review of Earth and Planetary Sciences, 38(1), 103-131, doi:10.1146/annurev-earth-040809-152537.

Tanioka, Y., Yudhicara, T. Kususose, S. Kathiroli, Y. Nishimura, S.-I. Iwasaki, and K. Satake (2006), Rupture process of the 2004 great Sumatra-Andaman earthquake estimated from tsunami waveforms, Earth, Planets and Space, 58(2), 203-209, doi:10.1186/BF03353379.

Teresa, J., M. Venugopal, D. Roy, R. M. Vasu, and R. Kanhirodan (2014), Diffraction tomography from intensity measurements: an evolutionary stochastic search to invert experimental data, J. Opt. Soc. Am. A, 31(5), 996-1006, doi:10.1364/JOSAA.31.000996.

Whittaker, J. M., A. Goncharov, S. E. Williams, R. D. Müller, and G. Leitchenkov (2013), Global sediment thickness data set updated for the Australian-Antarctic Southern Ocean, Geochemistry, Geophysics, Geosystems, 14(8), 3297-3305, doi:10.1002/ggge.20181.

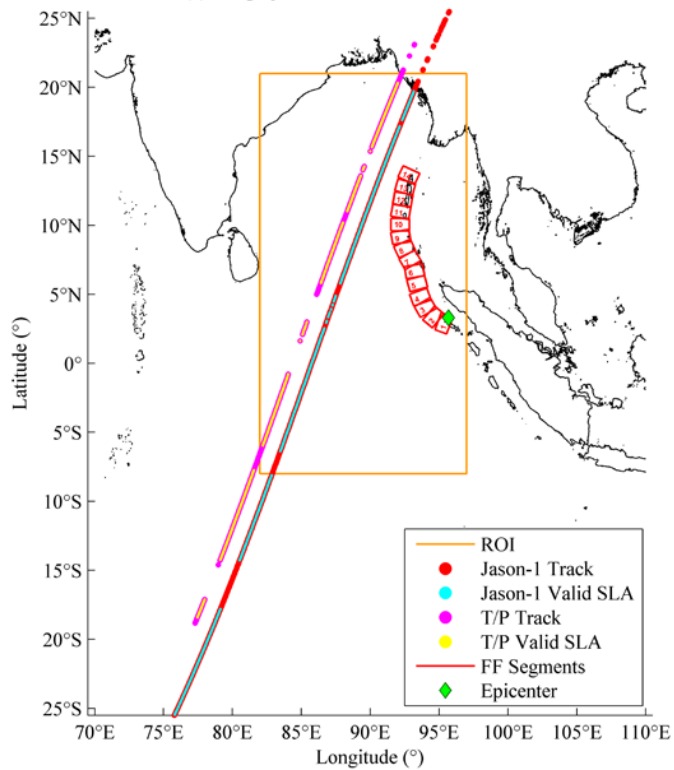
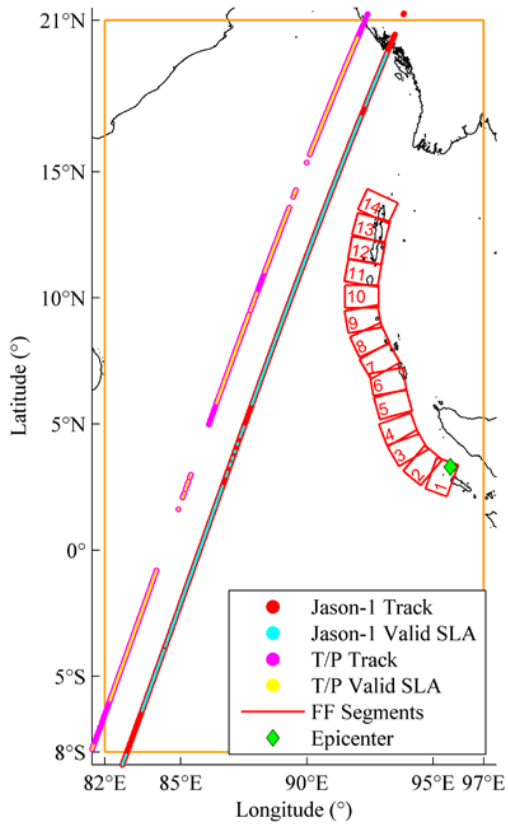
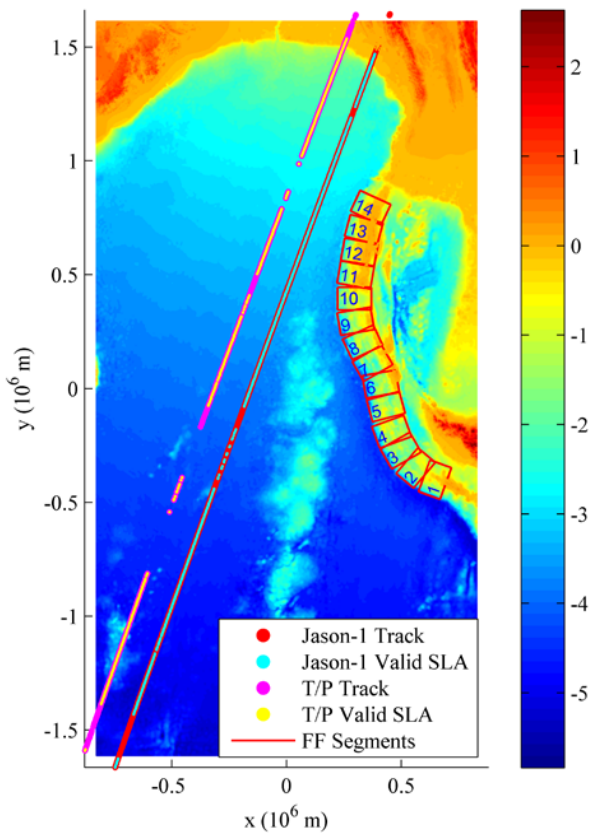

**Figure 1**. Geographical extent of 2004 S-A event. (a) Computational domain bounded by orange box. Satellite tracks and finite fault segmentation: (b) without and (c) over bathymetry (km).

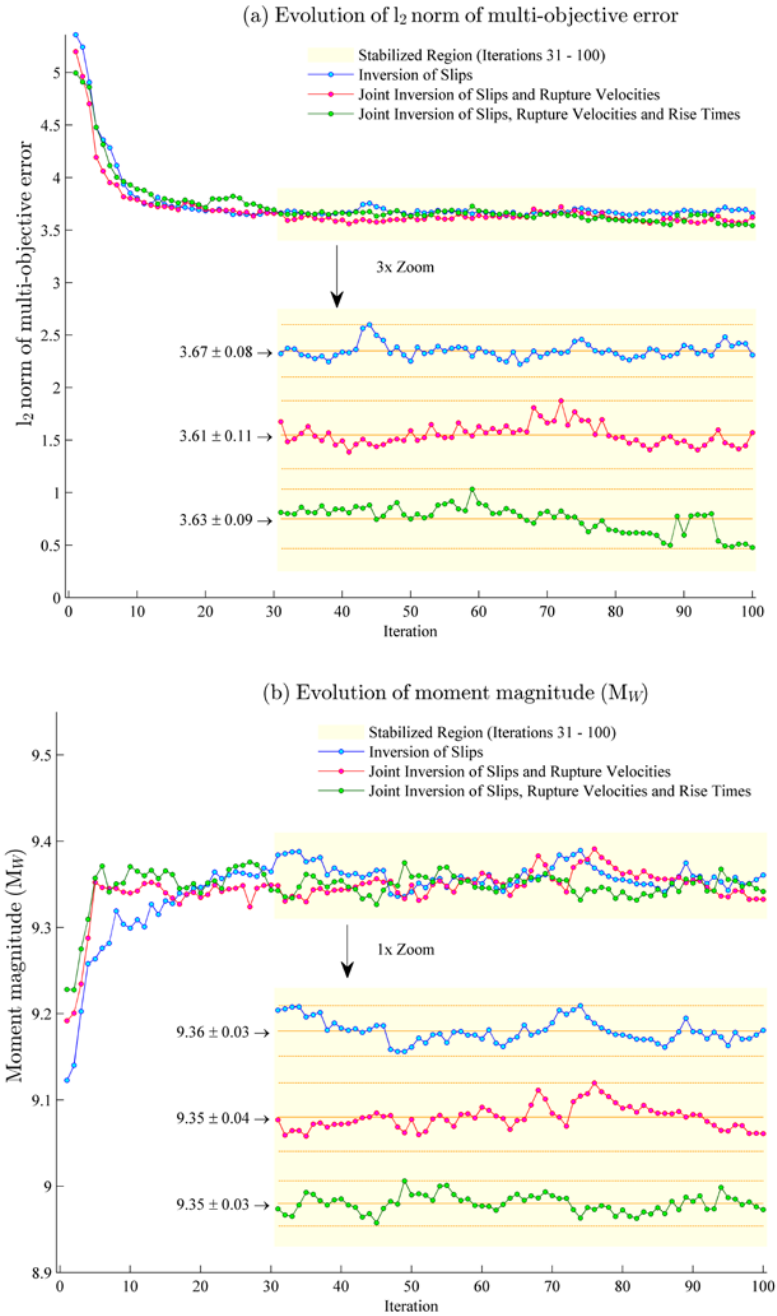

**Figure 2.** Evolution of relevant quantities; (a) $l_2$-norm of error and (b) moment magnitude.

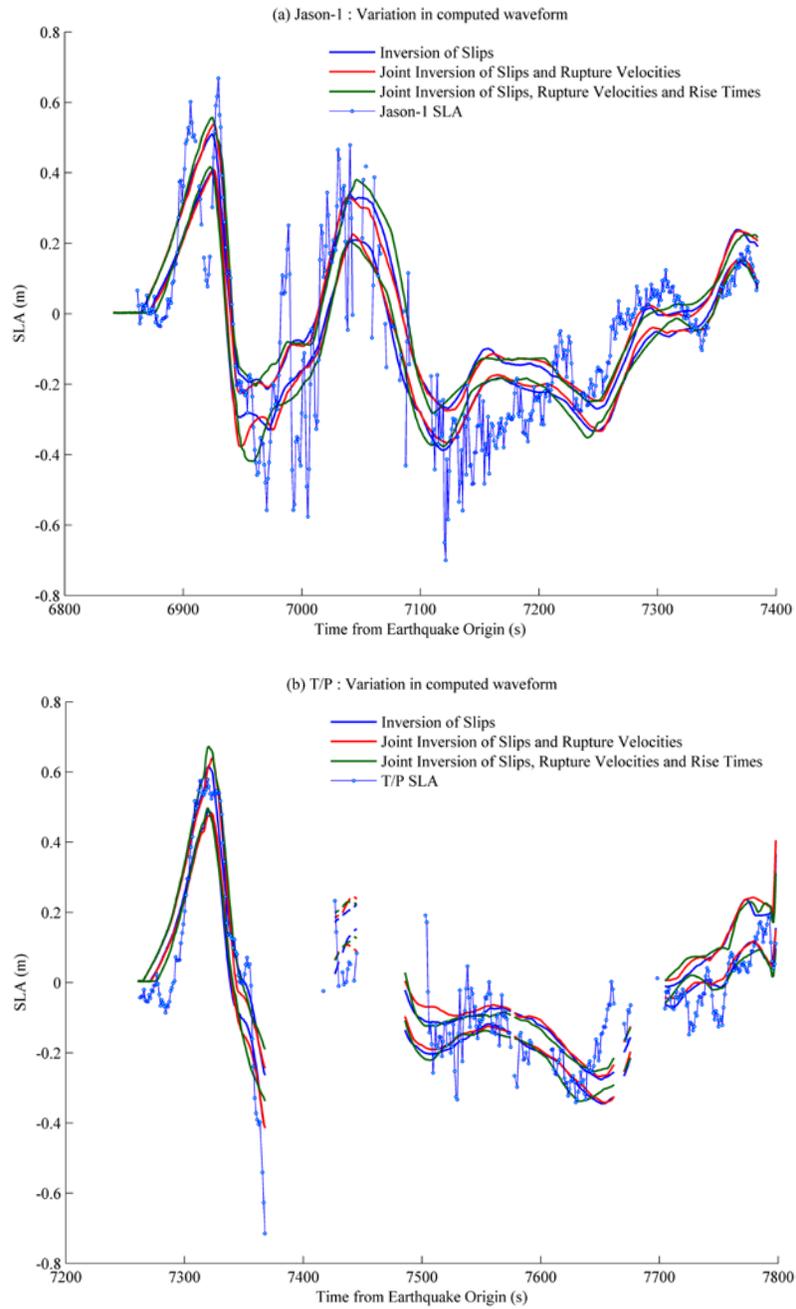

**Figure 3.** Variations in computed SLA during steady state. Their envelope boundaries for (a) Jason-1 and (b) T/P.

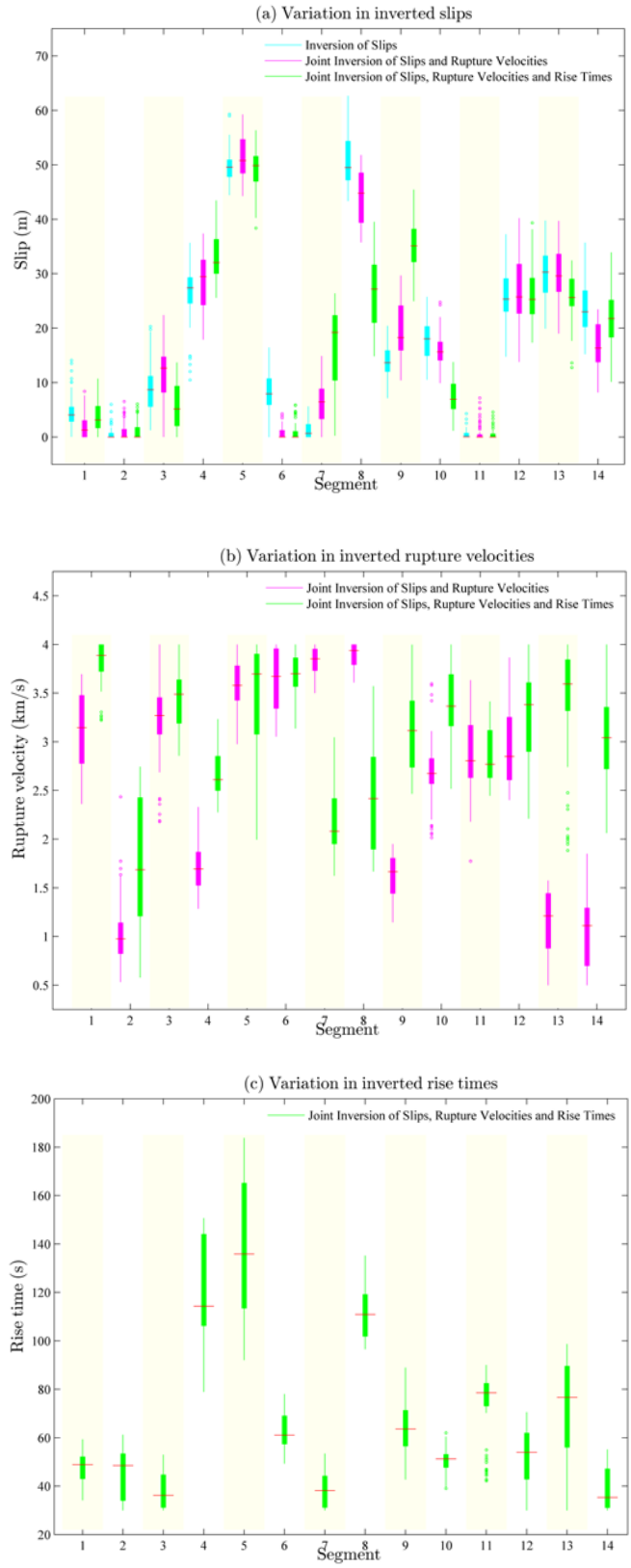

**Figure 4.** Variances in parameters. (a) Slips, (b) rupture velocities and (c) rise times.